\documentstyle[12pt]{article}

\begin{document}
\baselineskip 1.5 \baselineskip
\vspace{1cm}
\begin{center}
{\Large The generalized quantum statistics}
\end{center}

\vspace{1cm}

\begin{center}
 Won Young Hwang \footnote{ e-mail: wyhwang@correl1.snu.ac.kr } 
 ,Jeong-Young Ji \footnote{ e-mail: jyji@phya.snu.ac.kr}
 and Jongbae Hong \\
\vspace{0.1cm}
{\em  Department of Physics Education, \\ Seoul National
  University, \\ Seoul 151-742, Korea }
\end{center}

\vspace{0.8cm}
\begin{flushleft}
{\bf Abstract }
\end{flushleft}
The concept of wavefunction reduction should be 
introduced to standard quantum mechanics in any physical processes
where effective reduction of wavefunction occurs, as well as in the 
measurement processes. When the overlap is negligible, each particle 
obey Maxwell-Boltzmann even if the particles are in principle 
described by totally symmetrized wavefunction [P.R.Holland, The 
Quantum Theory of Motion, Cambridge University Press, 1993, p293].
We generalize the conjecture. That is, 
particles obey some generalized statistics that contains the quantum 
and classical statistics as special cases,
where the level of overlapping determines the statistics that particles
should obey among continuous generalized statistics. We present an
example consistent with the conjecture.
\vspace{0.8cm}
\begin{flushleft}
\end{flushleft}

\newpage
 
\section{Introduction}
It is generally known that it is the difference in the way of counting the 
number of accessible states that make quantum (Bose-Einstein or Fermi-Dirac)
statistics 
different from classical (Maxwell-Boltzmann) statistics \cite{ters,kipr,reif}:
For $N$ particles distributed among $M$ discrete states, the probability of a
set of occupancies, $\{n_i\}$, $i=1,2,...,M$ is simply propotional to the 
number of distinct configurations corresponding to $\{n_i\}$. If one counts
distinct configurations, one obtains Maxwell-Boltzmann (MB) statistics, 
with the probability
 \begin{equation}
  p\{n_i\}= \frac{N!}{M^N n_1! n_2! \cdot\cdot\cdot  n_M!}.
 \end{equation}
If one does not count distinct configurations or if one assume that each $\{n_i\}$
defines a unique configuration, one obtains Bose-Einstein (BE) statistics
 \begin{equation}
  p\{n_i\}= \frac{N!(M-1)!}{(N+M-1)!}.
 \end{equation}
If one further imposes the Pauli exclusion principle, Fermi-Dirac (FD) 
statistics is obtained.
Let us consider a simple but illustrating example.
When $N=M=2$,
in MB statistics,
 \begin{equation}
  p\{2,0\}=p\{0,2\}=\frac{1}{4}, \hspace{5mm}  p\{1,1\}=\frac{1}{2},
 \end{equation}
in BE statistics,
 \begin{equation}
  p\{2,0\}=p\{0,2\}=p\{1,1\}=\frac{1}{3},
 \end{equation}
in FD statistics,
 \begin{equation}
  p\{2,0\}=p\{0,2\}=0,  \hspace{5mm} p\{1,1\}=1,
 \end{equation}
In this case, the three statistics are distinct. In the case of $N=M=2$,
it is generally accepted that Eqs.(3)-(5) must be obeyed, or that there
is no intermediate case where 
 \begin{eqnarray}
  p\{2,0\}=p\{0,2\}=a,  \hspace{5mm} p\{1,1\}=1-2a, \\
  \mbox{where} \hspace{7mm} 0<a<\frac{1}{4}
  \hspace{7mm} \mbox{or} \hspace{7mm} \frac{1}{4}<a<\frac{1}{3} \nonumber
 \end{eqnarray}
The main purpose of this paper is to present some arguments and an example
which suggest that the intermediate case (Eq.(6)) will be obeyed by real
Bose or Fermi particles in the case where the overlapping between particles
are neither sufficient nor negligible.
One may wander if this intermediate case can be 
explained by the fact that BE and FD statistics approaches to MB statistics
in some (classical limit) cases where $1<<N<<M$. However, in our discussion,
$N$ and $M$ is fixed and thus the three statistics are distinct 
(Eqs.(3)-(5)) all the time. So this intermedate case is different from what
can be explained by the classical limit.   
In the next section, we will do some preliminary
discussions about how totally symmetrized (thus entangled) states can be 
dealt as some direct product (thus not entangled) state in some cases. 
In section3, we present discussion where we conjecture that real particle 
will obey some intermediate statistics between the quantum statistics and 
the classical one. In section4, we present an example consistent with the
conjecture.

\section{symmetrized and direct product state}
 Let us consider a 4-particle symmetrized state. Since the generalization
to n-particle ($n$ is a positive integer) and antisymmetrized case of the
following arguments is straitforward, we omit it for the sake of the space.
The 4-particle symmetrized state is
 \begin{eqnarray}
  |\Psi\rangle = \frac{1}{\sqrt{4!}} |\epsilon_{ijkl}|
                 \psi_i(x_1)\psi_j(x_2)\psi_k(x_3)\psi_l(x_4), \\
                 \mbox{where i,j,k,l = 1,2,3,4.}  \nonumber
 \end{eqnarray}
(The antisymmetrized state is obtained by replacing $|\epsilon_{ijkl}|$
by $\epsilon_{ijkl}$ in Eq.(7).)
As we see, the symmetrized state is an entangled state. So we cannot endow 
some subset of the 4-particle, say particle1, with some pure states. Hence, 
even if we want to describe only particle1, we must use the 4-particle 
symmetrized state in principle. However, single-particle pure states succesfully
works in some cases. This seemingly paradoxical fact is well known 
\cite{cohe,hol2} and resolved by the fact that as the overlap between 
particles are reduced the physical prediction of the 4-particle state 
(Eq.(7)) will approach to that of single-particle state. However, it should
be that the same thing can be said for all possible subsets of the 4
particles. That is, we must be able to explain, for example, how a state of
the form
\begin{equation}
  [\psi_1(x_1)\psi_2(x_2)+\psi_2(x_1)\psi_1(x_2)] 
  [\psi_3(x_3)\psi_4(x_4)+\psi_4(x_3)\psi_3(x_4)] 
\end{equation} 
succesfully works in some cases. This may also be explained in the same way
as above. In this section, we will describe from different perspectives how
the 4-particle symmetrized state (Eq.(7)) is reduced to some direct products
of subsystem states that are symmetrized in itself (Eq.(8)).
Now, let us assume that $\psi_1(x)$  and $\psi_2(x)$ 
($\psi_3(x)$  and $\psi_4(x)$) are peaked around $x \approx x_A$
($x \approx x_B$), while negligible except for the peaks.
We also assume that $x_A$ differs enough from $x_B$ so that 
$\psi_l(x)$ does not overlap with $\psi_m(x)$ ($l=1,2$ $m=3,4$).
Then, when $x_1 \approx x_A$,$x_2 \approx x_A$ and $x_3 \approx x_B$,
$x_4 \approx x_B$, the only nonnegligible terms in Eq.(7) is 
\begin{equation}
  [\psi_1(x_1)\psi_2(x_2)+\psi_2(x_1)\psi_1(x_2)] 
  [\psi_3(x_3)\psi_4(x_4)+\psi_4(x_3)\psi_3(x_4)]. 
\end{equation} 
Eq.(9) is just the state (Eq.(8)) to which the totally symmetrized state
(Eq.(7)) should be reduced.
In the guidence formula of the deBroglie-Bohm 
\cite{bohm,holl,boh2,bell} model, only the nonnegligible part 
play their roles. We can say that the total wavefunction (Eq.(7)) have
effectively reduced to Eq.(9) \cite{hol2,bell}. 
In this case, the 4-particle physical system behaves as if it is composed of two
2-particle entangled states each corresponding to 
$\psi_1(x_1)\psi_2(x_2)+\psi_2(x_1)\psi_1(x_2)$ and
$\psi_3(x_3)\psi_4(x_4)+\psi_4(x_3)\psi_3(x_4)$, although,
in principle, it is a 4-particle entangled state.
However, it is this effective reduction that seamlessly explains the reduction 
of wavefunction of measured system \cite{hol2,boh2,bell}. From the viewpoint of
the de Broglie-Bohm model, the effective reduction can occur in any physical
processes including the measurement process. Since the de Broglie-Bohm model
is, at least, equivalent to the standard quantum mechanics, we can say that 
{\it the concept of wavefunction reduction should be introduced to the standard
quantum mechanics in any physical
processes where the effective reduction of wavefunction occurs, not only in the 
measurement processes}. So in the case of the above 4-particle state, we should
say that the totally symmetrized state (Eq.(7)) has been reduced to Eq.(9), in 
the same way as we say that wavefunction of a system has been reduced to some 
wavefunction after measurements.

\section{ The implication of the wavefunction reduction on quantum statistics}
As explained in the previous section, it is the overlap between $\psi_i(x)$
constituting the totally symmetrized wavefunction that determines whether
we should use the totally symmetrized wavefunction (like Eq.(7)) or some 
partially symmetrized wavefunction (like Eq.(9)). Now, let us consider the 
counting ways of accesible states in the case of the previous example involving
Eqs.(7)-(9). Does the state of Eq.(9) should be regarded as distinct from the 
state 
\begin{equation}
  [\psi_3(x_1)\psi_4(x_2)+\psi_4(x_1)\psi_3(x_2)] 
  [\psi_1(x_3)\psi_2(x_4)+\psi_2(x_3)\psi_1(x_4)] ?
\end{equation} 
There may be two answers to this question: firstly, we can say that they
(Eqs.(9)and(10)) should not be regarded as distinct states since the 4 particles
are in principle in the totally symmetrized state (Eq.(7)). Secondly, we can 
say that they should be regarded as distinct ones since they have been reduced to
one of the two states. If the latter is correct, the group of the particle1
and particle2 (or particle3 and particle4), as a whole, will obey the MB statistics
while within the group each individual particle will obey BE statistics. 
This is consistent with 
the general assumption of canonical distribution. So we can say that the latter 
is correct. Thus we can also say that {\it the degree of overlap determines the 
counting ways and ,as a result, the statistics.} Thus, when each $\psi_i(x)$ is 
overlapping sufficiently (negligibly) with each other, the particles obey 
quantum (classical) statistics. Then, we can pose the following question. Which
statistics should particles obey when the overlapping is neither sufficient nor
negligible? In this intermediate case, it seems that the particles will obey
some intermediate statistics between quantum and classical statistics, as the 
one of Eq.(6) in the case of $M=N=2$. One may wander how this intermediate 
statistics will be possible in spite of the fact that there is no intermediate
particles between quantum and classical ones. Here we underline the fact that
it is not the symmetry of wavefunctions but the properties of (many particle
time-dependent) Schrodinger equation's solutions that directly determine the
statistics of particles. In other words, the quantum equipartition principle is
mere consequences of Schrodinger equation, as if the classical equipartition
principle 
is consequences of Newtonian equantion. Summarizing this section, we say that
{\it particles obey some generalized
statistics that contains the quantum and classical statistics as special cases.
The level of overlapping determines the statistics that
particles should obey among
continuous generalized statistics.}

\section{ensemble of particle pairs tunneling through a potential barrier}
Now, let us consider an
ensemble of pairs of (identicle) particles that tunnel through a potential barrier.
We assume that the probability of a particle's reflecting on (tunneling through)
the barrier is $\frac{1}{2}$. Then, what is the probability $p\{0,2\}$ 
($p\{2,0\}$) for both particles to tunnel through (to reflect on) the barrier?
What is the probability $p\{1,1\}$ that one particle tunnels and the other
particle reflects? There are two ways of calculating 
the probability \cite{garu}: with MB statistics-like ways,
\begin{equation}
  p\{2,0\}=p\{0,2\}=\frac{1}{4}, \hspace{5mm}  p\{1,1\}=\frac{1}{2},
\end{equation}
with BE statistics-like ways 
\begin{equation}
  p\{2,0\}=p\{0,2\}=p\{1,1\}=\frac{1}{3}.
\end{equation}
 (It is assumed in Ref.\cite{garu} that the latter is correct.) 
Applying the assumption about the generalized statistics of the previous section,
we will get the following result. When each member of particle pairs is overlapping
sufficiently (negligibly) with each other, the ensemble should be described by
2-particle symmetrized state of the form
\begin{equation}
  \Psi(x_1,x_2,t)
  =\frac{1}{\sqrt{2}}[\psi_A(x_1,t)\psi_B(x_2,t)+\psi_B(x_1,t)\psi_A(x_2,t)]
\end{equation}
(by a direct product of two 
single-particle states) and the ensemble will obey Eq.(12) obtained by 
BE statistics-like ways (Eq.(11) obtained by MB statistics-like ways). We 
note that Eq.(13) is a solution of 2-particle Schrodinger equation if each
constituting single-particle wavefunction $\psi_A(x,t)$ and $\psi_B(x,t)$ is 
a solution of Schrodinger equation.
Let us assume that each wavefunction $\psi_A(x_1,t)$ and $\psi_B(x_2,t)$ is
traveling Gaussian wavepackets that tunnels through a potential barrier with 
a probability $\frac{1}{2}$.
Wavepackets evolutions of this type for single particle and one-dimensional
case are well studied \cite{leav}.
The wavepacket starts to move to right from a certain point to the barrier.
Later, the wavepacket collides the barrier and splits to two parts,
one of which, after tunneling the barrier, continues to go to right, 
while the other one of which, after reflecting on the barrier, goes to left. 
Now, in order to test the conjecture of the previous section in the case of
this example, we estimate the statistics by analyzing the value
\begin{eqnarray}
  |\Psi(x_1,x_2,t)|^2
  =\frac{1}{2}\{ |\psi_A(x_1,t)|^2 |\psi_B(x_2,t)|^2
                +|\psi_B(x_1,t)|^2 |\psi_A(x_2,t)|^2 \nonumber\\
             +2Re[\psi_A(x_1,t)^* \psi_B(x_1,t)
                 (\psi_A(x_2,t)^* \psi_B(x_2,t))^*] \}.
\end{eqnarray}
 $|\Psi(x_1,x_2,t)|^2$ is the probability density of finding particle1 at 
$x_1$, particle2 at $x_2$ at a time $t$. Note that we are now
working in the de Broglie-Bohm model. So there is no problem in the 
interpretation like this. Since the de Broglie-Bohm model is at least 
equivalent to the standard quantum mechanics, some results obtained using
the Broglie-Bohm model will be as correct as the standard quantum mechanics.
So we obtain that 
\begin{eqnarray}
 p\{0,2\}&=&\int_0^\infty    \int_0^\infty    |\Psi(x_1,x_2,t)|^2 dx_1 dx_2, \\
 p\{2,0\}&=&\int_{-\infty}^0 \int_{-\infty}^0 |\Psi(x_1,x_2,t)|^2 dx_1 dx_2, \\
 p\{1,1\}&=&\int_0^\infty    \int_{-\infty}^0 |\Psi(x_1,x_2,t)|^2 dx_1 dx_2 
           +\int_{-\infty}^0 \int_0^\infty    |\Psi(x_1,x_2,t)|^2 dx_1 dx_2, \\
         && \mbox{where} \hspace{3mm}t \hspace{3mm} \mbox{is sufficently large}.   \nonumber
\end{eqnarray}
When $\psi_A(x,t)$ and $\psi_B(x,t)$ does not overlap, the last term of Eq.(14)
, on average, vanishes. 
In this case, we obtain 
\begin{equation}
p\{2,0\} \approx \frac{1}{4}, \hspace{5mm} p\{0,2\} \approx \frac{1}{4}
\hspace{5mm}  p\{1,1\} \approx \frac{1}{2}.
\end{equation}
However, when $\psi_A(x,t)$ and $\psi_B(x,t)$ overlap much, the last term of 
Eq.(14) contribute to the integrals at some cases: let us assume that 
\begin{equation}
 \psi_A(x,t) \approx \psi_B(x,t)
\end{equation}
and 
\begin{equation}
 x_1 \approx x_2.
\end{equation}
Then the last term of Eq.(14) contributes to the integrals as much as remaining
terms of the righthandside of Eq.(14). However, the condition (Eq.(20)) is 
satisfied in the case of Eqs.(15) and (16) but not in the case of Eq.(17).
Thus, $p\{2,0\}$ and $p\{0,2\}$ become multiplied by a factor $2$, approximately.
Therefore,
\begin{equation}
  p\{2,0\} \approx \frac{1}{3} \hspace{5mm}
  p\{0,2\} \approx \frac{1}{3} \hspace{5mm}
  p\{1,1\} \approx \frac{1}{3}.
\end{equation}
However, when the overlap is intermediate,
we will obtain some results between Eq.(18)
and (21), as we have expected. 
Summarizing these, we can say that in accordance with the level of
overlapping between $\psi_A(x,t)$ and $\psi_B(x,t)$, the (Bose) particles 
will obey some statistics
\begin{eqnarray}
  p\{2,0\}=p\{0,2\}=a,  \hspace{5mm} p\{1,1\}=1-2a, \\
  \mbox{where}  \hspace{7mm} \frac{1}{4}<a<\frac{1}{3}. \nonumber
\end{eqnarray}
The generalization to Fermi particles can be done by replacing the $+$ sign
by the $-$ sign in Eq.(13). Then,
\begin{eqnarray}
  p\{2,0\}=p\{0,2\}=a,  \hspace{5mm} p\{1,1\}=1-2a, \\
  \mbox{where} \hspace{7mm} 0<a<\frac{1}{4}. \nonumber
\end{eqnarray}
However, since the example of this section do not describe an equilibrium
processes, we can not say this example verifies the conjecture of
the previous section although this example is consistent with the conjecture.

\section{discussion and conclusion}
It may be that in most cases the intermediate (or generalized) statistics
is not needed in describing physical phenomena. However, the generalized 
statistics needs to be searched for.  
To summarize, the concept of wavefunction reduction should be 
introduced to standard quantum mechanics in any physical processes
where effective reduction of wavefunction occurs, as well as in the 
measurement processes. With this, we have made a conjecture that 
particles obey some generalized statistics that contains the quantum 
and classical statistics as special cases,
where the level of overlapping determines the statistics that particles
should obey among continuous generalized statistics. We present an
example consistent with the conjecture.

\section*{Acknowledgments} 
We are very grateful for the support by the Korea Science 
and Engineering Foundation (KOSEF).

\end{document}